\documentstyle[twocolumn,floats,aps,prl,epsf]{revtex} 
\draft
\preprint{{\Large \sl ROUGH DRAFT}}
\title{\boldmath Glassy Spin Dynamics in Non-Fermi-Liquid 
UCu$_{5-x}$Pd$_x$, $x = 1.0$ and 1.5}
\author{D.~E. MacLaughlin,$^1$ O. O. Bernal,$^2$ R.~H. Heffner,$^3$ 
G.~J. Nieuwenhuys,$^4$ M.~S. Rose,$^1$ \\ J.~E. Sonier,$^3$\cite{SFU} 
B. Andraka,$^5$ R. Chau,$^6$\cite{LLL} and M.~B. Maple$^6$}
\address{$^1$Department of Physics, University of California, 
Riverside, California 92521-0413 \\
$^2$Department of Physics, California State University, Los Angeles, 
California 90032 \\
$^3$MS K764, Los Alamos National Laboratory, Los Alamos, New Mexico 
87545 \\
$^4$Kamerlingh Onnes Laboratory, Leiden University, 2300 RA Leiden, 
the Netherlands \\
$^5$Department of Physics, University of Florida, Gainesville, 
Florida 32611 \\
$^6$Department of Physics, University of California, San Diego, La 
Jolla, California 92093 \\
\rm\small version date June 14, 2001		
}

\address{\parbox{137mm}{\bigskip\rm\small	
Local {\em f\/}-electron spin dynamics in the non-Fermi-liquid 
heavy-fermion alloys UCu$_{5-x}$Pd$_x$, $x = 1.0$ and 1.5, have been 
studied using muon spin-lattice relaxation. The sample-averaged 
asymmetry function~$\overline{G}(t)$ indicates strongly inhomogeneous 
spin fluctuations, and exhibits the scaling $\overline{G}(t,H) = 
\overline{G}(t/H^\gamma)$ expected from glassy dynamics. At 0.05~K 
$\gamma(x{=}1.0) = 0.35{\pm}0.1$, but $\gamma(x{=}1.5) = 0.7{\pm}0.1$. 
This is in contrast to inelastic neutron scattering results, which 
yield $\gamma = 0.33$ for both concentrations. There is no sign of 
static ${\rm magnetism} \gtrsim 10^{-3}~\mu_B$/U ion in either material  
above 0.05~K\@. Our results strongly suggest that both alloys are 
quantum spin glasses.
\\[6pt] PACS numbers: 71.27.+a, 75.30.Mb, 76.60.Cq.}}
\begin{document} \maketitle			

Magnetic resonance \cite{BMLA95,MBL96,BMAF96,AGVO99} and other 
\cite{BMHC98,dACDD98} experiments have demonstrated the importance of 
structural disorder in the breakdown of Landau's Fermi-liquid theory 
in certain {\em f\/}-electron intermetallic compounds and alloys. 
Disorder-driven mechanisms have been considered for the 
non-Fermi-liquid (NFL) properties of some of these 
systems~\cite{MDK96,CNCJ98}, and it is natural to consider the 
possibility of extremely disordered or ``glassy'' behavior. On 
theoretical and experimental grounds it is known that glassy dynamics 
lead to long-time correlations with distinct signatures as the freezing 
or ``glass'' temperature~$T_g$ is approached from above~\cite{KMCL96}. 
In a spin glass the spin autocorrelation function~$q(t) = \langle 
{\bf S}_i(t)${\boldmath $\cdot$}${\bf S}_i(0) \rangle$ is theoretically 
predicted to exhibit power-law ($q(t) = ct^{-\alpha}$) or 
``stretched-exponential'' ($q(t) = c\exp[-(\lambda t)^\beta]$) 
behavior~\cite{PSAA84}. Power-law correlation has been found in 
spin-glass AgMn using muon spin relaxation ($\mu$SR)~\cite{KMCL96}. 

This Letter describes evidence from $\mu$SR experiments that spin 
correlations in the NFL alloys~UCu$_{5-x}$Pd$_x$, $x = 1.0$ and 1.5, 
are indicative of glassy spin dynamics. The sample-averaged muon 
relaxation function (asymmetry)~$\overline{G}(t,H)$ is strongly 
sub-exponential, indicating a quenched inhomogeneous distribution of 
relaxation rates, and obeys the time-field scaling 
relation~$\overline{G}(t,H) = \overline{G}(t/H^\gamma)$ for applied 
magnetic field~$H$ between $\sim$15~Oe and $\sim$1 kOe. The field 
dependence corresponds to a measurement of the Fourier transform of 
$q(t)$ over the frequency range~$\gamma_\mu H/2\pi \approx 
200$~kHz--14~MHz, where $\gamma_\mu = 2\pi \times 13.55$~kHz/Oe is the 
muon gyromagnetic ratio. Power-law behavior of $q(t)$ is implied by the 
observation~$\gamma < 1$~\cite{KMCL96}, and also by the 
temperature-frequency scaling found in the inelastic neutron scattering 
(INS) cross section~\cite{AORL95}, although the possible connection with 
glassy dynamics has to our knowledge not been previously noted. The 
present measurements extend by three orders of magnitude the frequency 
range over which power-law correlations are observed in 
UCu$_{5-x}$Pd$_x$. Zero-field $\mu$SR above 0.05~K shows no sign of 
static magnetism or spin freezing in UCu$_4$Pd; this together with the 
glassy scaling points to some form of quantum spin-glass 
behavior~\cite{Sach98,GrRo99}. 

In $\mu$SR experiments spin-polarized positive muons are implanted into 
the sample, and the subsequent time evolution of the muon polarization 
is monitored by detecting the asymmetric distribution of positrons from 
the muon decay~\cite{Sche85}. Muon relaxation in a magnetic field 
applied parallel to the muon spin direction (longitudinal field) is due 
to thermally-excited {\em f\/}-electron spin fluctuations that couple 
to the muons. A muon at a given site experiences a time-varying local 
field~${\bf H}_{\rm loc}(t)$ due to fluctuations of neighboring {\em 
f\/} moments. Following Keren {\em et al.\/}\cite{KMCL96}, under 
motionally narrowed conditions the local muon asymmetry~$G(t,H)$ 
relaxes exponentially:
\begin{equation}
G(t,H) = \exp\left[ -2\Delta^2 \tau_c(H)t \right] \,,
\label{eq:relax}
\end{equation}
where $\Delta^2 = \gamma_\mu^2\langle |{\bf H}_{\rm loc}|^2\rangle$ is 
the time-averaged mean-square coupling constant in frequency units, and 
the local correlation time~$\tau_c(H)$ is given by
\begin{equation}
\tau_c(H) = \int_0^\infty\!\! dt\, q(t) \cos(\gamma_\mu Ht) = cu_c(H) 
\,.
\label{eq:tauc}
\end{equation}
We consider $\Delta$ and the prefactor~$c$ but not the functional form 
of $u_c(H)$ to vary from site to site in a disordered material. Then 
the sample-averaged asymmetry~$\overline{G}(t,H)$ is given by
\begin{equation}
\overline{G}(t,H) = \int\!\!\int d\Delta dc\, \rho(\Delta,c) \exp\left[ 
-2\Delta^2 cu_c(H)t \right] \,,
\label{eq:Gav}
\end{equation}
where $\rho(\Delta,c)$ is the joint distribution function for $\Delta$ 
and $c$. It can be seen that the field and time dependence enter 
Eq.~(\ref{eq:Gav}) in the combination~$u_c(H)t$. This means that 
$\overline{G}(t,H)$ scales as this combination independently of the 
form of $\rho(\Delta,c)$. 

For both the power-law and stretched-exponential forms of $q(t)$ this 
time-field scaling results in~\cite{KMCL96}
\begin{equation}
\overline{G}(t,H) = \overline{G}(t/H^\gamma)
\label{eq:scaling}
\end{equation}
after Fourier transforming $q(t)$. Here $\gamma = 1 - \alpha < 1$ for 
power-law correlation, and $\gamma = 1 + \beta > 1$ for 
stretched-exponential correlation as long as the muon Larmor 
frequency~$\omega_\mu = \gamma_\mu H$ is much greater than 
$\lambda$~\cite{KMCL96}. If Eq.~(\ref{eq:scaling}) is obeyed a plot of 
$\overline{G}(t,H)$ versus $t/H^\gamma$ will be universal for the 
correct choice of $\gamma$. The sign of $\gamma - 1$ distinguishes 
between power-law and stretched-exponential correlations.

\begin{figure}[t]
\begin{center}
\epsfxsize 2.95in \leavevmode
\epsfbox{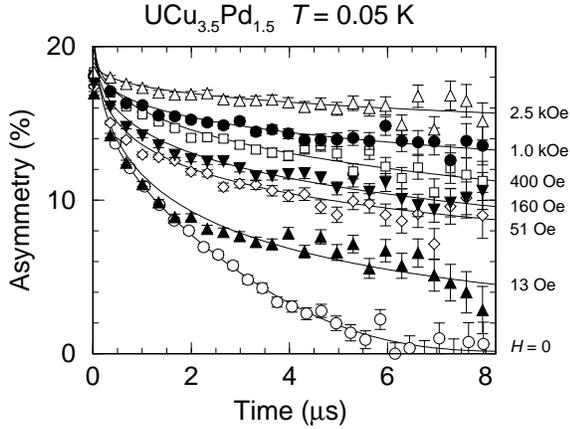}
\end{center}
\caption{Field dependence of sample-averaged muon asymmetry relaxation 
function~$\overline{G}(t)$ in UCu$_{3.5}$Pd$_{1.5}$, $T = 0.05$~K\@. 
Curves: fits as described in text.}
\label{fig:UC3Pdasy}
\end{figure}
Samples of UCu$_4$Pd and UCu$_{3.5}$Pd$_{1.5}$ were prepared as 
described previously~\cite{AnSt93}. Arc-melted ingots were crushed into 
powder under acetone. The powder was passed through a 90 micron sieve, 
and pressed with GE 7031 varnish into pellets 13~mm ${\rm dia.} \times 
{\sim}1$~mm thickness. $\mu$SR data were taken at the LTF facility of 
the Paul Scherrer Institute, Villigen, Switzerland, for temperatures 
between 0.05 and 1.1~K and for magnetic fields between zero and 10~kOe 
applied in the direction of the muon spin polarization.

\begin{figure}[t]
\begin{center}
\epsfxsize 2.95in \leavevmode
\epsfbox{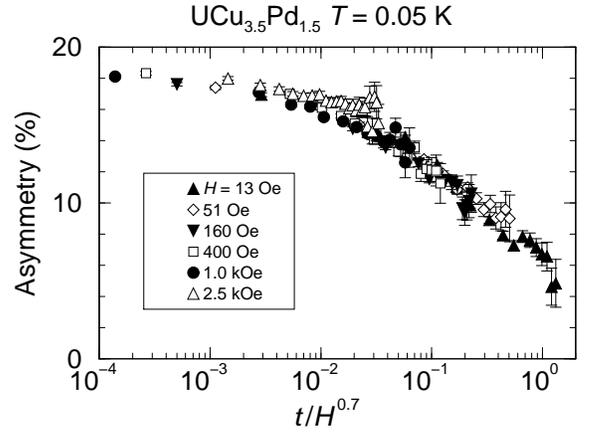}
\end{center}
\caption{Data of Fig.~\protect\ref{fig:UC3Pdasy} plotted versus the 
scaling variable~$t/H^{0.7}$.} 
\label{fig:UC3Pscal}
\end{figure}

Figure \ref{fig:UC3Pdasy} shows $\overline{G}(t)$ in 
UCu$_{3.5}$Pd$_{1.5}$ for $T = 0.05$~K and values of applied field~$H$ 
between 13~Oe and 2.5~kOe. The relaxation slows with increasing field. 
For low enough fields we expect the field dependence to be due to the 
change of $\omega_\mu$ rather than an effect of field on $q(t)$; a 
breakdown of scaling would occur for high fields where this ceases 
to be true. The same asymmetry data are plotted in 
Fig.~\ref{fig:UC3Pscal} as a function of the scaling 
variable~$t/H^\gamma$. For $\gamma = 0.7 \pm 0.1$ the data scale well 
over more than three orders of magnitude in $t/H^\gamma$ and for all 
fields except 2.5~kOe. Fields $\mu_BH \gtrsim k_BT$ would be expected 
to affect the spin dynamics, and indeed the static susceptibility of 
UCu$_4$Pd is suppressed by fields $\sim$1 kOe below $\sim$0.5 K 
(Vollmer {\em et al.\/}, Ref.~\cite{SSHK98}). The scaling 
exponent~$\gamma$ is less than 1, implying that $q(t)$ is well 
approximated by a power law (or a cutoff power law~\cite{KMCL96}) 
rather than a stretched-exponential or exponential. From our data 
$q(t) \approx ct^{-0.3 \pm 0.1}$. We note again that no specific form 
for the muon asymmetry function has been assumed.

\begin{figure}[t]
\begin{center}
\epsfxsize 2.95in \leavevmode
\epsfbox{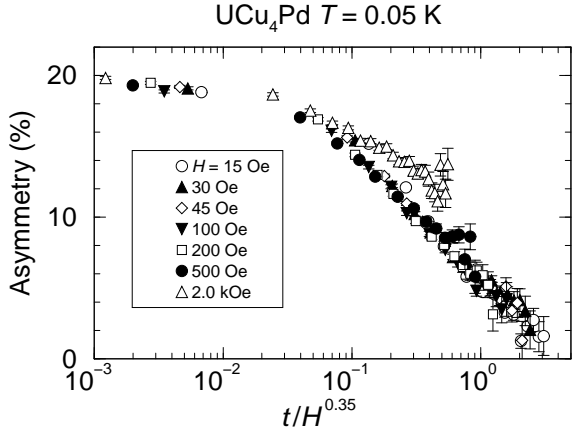}
\end{center}
\caption{Scaling plot of $\overline{G}(t,H)$ vs $t/H^{0.35}$ for 
UCu$_4$Pd, $T = 0.05$~K\@.} 
\label{fig:UC4Pscal}
\end{figure}

A scaling plot is given in Fig.~\ref{fig:UC4Pscal} for UCu$_4$Pd, $T = 
0.05$ K\@. Here the scaling exponent~$\gamma = 0.35 \pm 0.1$ is 
significantly smaller than in UCu$_{3.5}$Pd$_{1.5}$. Scaling again 
breaks down for high enough fields; the data for 2~kOe clearly fall 
above the low-field scaling curve. Data taken in UCu$_4$Pd at $T = 
0.5$~K (not shown) scale with the same exponent. This is in contrast to 
the scaling behavior of spin-glass 
Ag$_{0.995}$Mn$_{0.005}$~\cite{KMCL96}, where $\gamma$ varies strongly 
with temperature as the glass temperature~$T_g = 2.95$~K is approached 
from above. 

The fluctuation-dissipation theorem \cite{FDT} relates $\tau_c(H)$ to the 
imaginary component~$\chi''(\omega)$ of the local ($q$-inde\-pendent) 
{\em f\/}-electron dynamic susceptibility:
\begin{equation}
\tau_c(H) \approx \frac{k_BT}{\mu_B^2} \left( 
\frac{\chi''(\omega)}{\omega} \right) 
\label{eq:FDthm}
\end{equation}
for $\hbar\omega \ll k_BT$. INS experiments~\cite{AORL95} show 
scaling of the sample-averaged $\overline{\chi}''(\omega,T)$ as 
$\omega^{-\gamma} F(\hbar\omega/k_B T)$, with $\gamma = 0.33$ and $F(x) 
= \tanh(x/1.2)$ for both UCu$_4$Pd and UCu$_{3.5}$Pd$_{1.5}$. Using 
this form $\overline{\tau}_c(H)$ obtained from Eq.~(\ref{eq:FDthm}) is 
independent of $T$ and proportional to $H^{-\gamma}$; the latter is in 
accord with the $\mu$SR scaling. The INS value of $\gamma$ agrees with 
$\mu$SR data for UCu$_4$Pd ($\gamma = 0.35$) but not for 
UCu$_{3.5}$Pd$_{1.5}$ ($\gamma = 0.7$), suggesting that in the latter 
sample the behavior of $\tau_c$ changes in the unexplored region 
between the highest muon frequencies ($\sim$10~MHz) and the lowest INS 
frequencies ($\sim$10~GHz). Clearly high-resolution INS studies 
(neutron spin echo or backscattering) would be desirable between 
these frequencies. 

To go further one must fit the $\mu$SR data to an appropriate 
functional form for the asymmetry. We have chosen the 
stretched-exponential
\begin{equation}
\overline{G}(t) = \exp[-(\Lambda t)^K]\,, 
\label{eq:stretch}
\end{equation}
where $K < 1$ gives sub-exponential relaxation corresponding to a 
distribution of relaxation rates. This function is purely empirical. It 
is used because it characterizes an {\em a priori\/} unknown 
relaxation-rate distribution, and because the rate~$\Lambda$ conforms 
with a general definition of a characteristic rate by the 
time~$1/\Lambda$ where $\overline{G}(t)$ decays to $1/e$ of its initial 
value. Equation~(\ref{eq:stretch}) fits the data in applied field to 
within the statistical error.

For $H = 0$ the data were fit to the product of Eq.~(\ref{eq:stretch}) 
and the zero-field Kubo-Toyabe (K-T) function~\cite{HUIN79} 
characteristic of static relaxation by nuclear dipolar fields at muon 
sites. This form is expected when the muon local field has both static 
nuclear dipolar and dynamic {\em f\/}-moment contributions. A nuclear 
dipolar field $\sim$2.3~Oe was measured in both alloys for $T 
\gg 1$ K, where the contribution of U-moment fluctuations to the
zero-field muon relaxation rate vanishes. Nonzero values of $H$ were 
chosen large enough to ``decouple'' the muon relaxation from the 
nuclear dipolar field~\cite{HUIN79} leaving only the dynamic U-moment 
contribution, so relaxation data for these fields were fit to 
Eq.~(\ref{eq:stretch}) without the K-T function. Curves giving these 
fits are plotted in Fig.~\ref{fig:UC3Pdasy}.

\begin{figure}[t]
\begin{center}
\epsfxsize 2.95in \leavevmode
\epsfbox{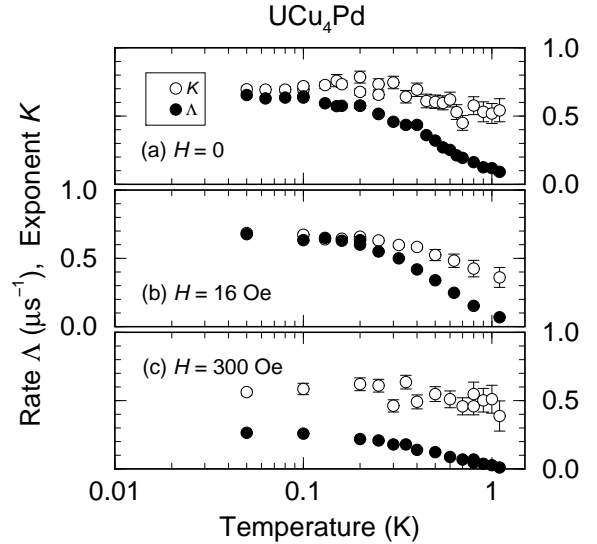}
\end{center}
\caption{Temperature dependence of muon stretched-expo\-nential 
relaxation rate~$\Lambda$ (filled circles) and exponent~$K$ (open 
circles) in UCu$_4$Pd. (a)~Applied longitudinal field~$H = 0$. (b)~$H 
= 16$~Oe. (c)~$H = 300$~Oe.}
\label{fig:UCu4Pd_temp}
\end{figure}

Figure \ref{fig:UCu4Pd_temp} gives $\Lambda(T)$ and $K(T)$ for 
UCu$_4$Pd at three values of $H$.
Below 1.1~K $\Lambda$ increases slowly and saturates to a constant 
below 0.1--0.2~K\@. As noted previously a temperature dependence of 
$\Lambda(T)$ is not in agreement with the temperature-independent 
relaxation predicted from INS scaling. The exponent~$K$ is 
approximately 0.7 at 0.05~K, indicative of a broad distribution of 
relaxation rates~\cite{KMCL96}, and decreases slightly with increasing 
temperature. Similar behavior is exhibited by $\Lambda(T)$ and $K(T)$ 
in UCu$_{3.5}$Pd$_{1.5}$ (data not shown), with rates slower than in 
UCu$_4$Pd by $\sim$30\% at low fields and $\sim$100\% at 100--300~Oe 
due to the larger scaling exponent.

No anomaly is found in the zero-field data at temperatures 
$\sim$0.1--0.2~K, where specific heat and ac susceptibility (in 
different samples) suggest spin-glass-like freezing~\cite{SSHK98}. 
The muon--{\em f\/}-moment coupling in UCu$_{5-x}$Pd$_x$ is 
predominantly dipolar~\cite{BMAF96} with a coupling field~$0.55 \pm 
0.05~{\rm kOe}/\mu_B$. Randomly-frozen moments of the order of 
1~$\mu_B$/U ion would result in a muon relaxation ${\rm rate} \sim 
50~\mu{\rm s}^{-1}$, two orders of magnitude larger than the observed 
rate. This result places an upper bound of ${\sim}10^{-3}~\mu_B$/U ion 
on any frozen moment in UCu$_4$Pd or UCu$_{3.5}$Pd$_{1.5}$. We believe 
that the discrepancy with the results of Refs.~\cite{SSHK98} results 
from differences in annealing conditions; $\mu$SR experiments to 
explore this question are currently underway. It is noteworthy that 
the saturation of $\Lambda$ and $K$ occurs in the same temperature 
range as the reported spin-glass-like freezing~\cite{SSHK98}.  

Our results can be compared with existing theories of disorder-driven 
NFL behavior. Preliminary analysis has indicated that the order of 
magnitude of the experimental rates at these very low temperatures 
cannot be directly accounted for by the simple single-ion Kondo 
disorder model of NFL behavior~\cite{MacL00}. In an early version of the 
``Griffiths-phase'' theory~\cite{CNCJ98}, which treats the effect of 
{\em f\/}-moment clustering, there is no dissipation in the {\em 
f\/}-electron spin dynamics, and the local cluster dynamic 
susceptibility is sharply resonant at a distributed characteristic 
tunneling energy~$E$:
\begin{equation}
\chi''(\omega,E) \propto \delta(\omega - E)\,\tanh(E/2T)\,.
\end{equation}
Together with the distribution function~$P(E) \propto E^{-1+\lambda}$, 
where $\lambda < 1$ is a nonuniversal scaling exponent, this 
immediately yields
\begin{equation}
\overline{\chi}''(\omega) = \int dE\, P(E)\chi''(\omega,E) \propto 
\omega^{-1+\lambda}
\tanh(\omega/2T) \,,
\label{eq:Grif}
\end{equation} 
in agreement with the INS data for $\lambda \approx 0.7$ and 
reminiscent of our $\mu$SR results. 

But the observed time-field scaling of $\overline{G}(t,H)$ demonstrates 
that it is the local $\chi''(\omega)$ itself, not just the average 
$\overline{\chi}''(\omega)$, which scales as $\omega^{-\gamma}$ [cf.\ 
Eqs.~(\ref{eq:tauc}--\ref{eq:FDthm})]. This is not a property of the 
Griffiths-phase model, in which the scaling is found only after the 
average has been taken, and only for a sharply resonant (and 
nonscaling) form of $\chi''(\omega,E)$. A recent form of this 
theory~\cite{CNJ00} considers dissipative effects, which broaden 
$\chi''(\omega,E)$ but do not give it a scaling form. Furthermore, if 
the width of $\chi''(\omega)$ is much greater than $\omega$ it is not 
hard to show that $\overline{\chi}''(\omega)$ no longer follows 
$P(\omega{=}E)$ so that this mechanism for scaling of 
$\overline{\chi}''(\omega)$ is lost. It is also difficult to see how 
the dynamic susceptibility of Ref.~\cite{CNJ00} would yield the 
observed temperature dependence of $\Lambda(\omega_\mu,T)$. Thus the 
Griffiths-phase theory does not seem to account for our results for 
a number of reasons.

Muon relaxation in UCu$_{5-x}$Pd$_x$ clearly indicates glassy dynamics, 
and the increase of $\Lambda(T)$ at low temperatures 
(Fig.~\ref{fig:UCu4Pd_temp}) suggests a low-temperature critical point. 
The increase in $K(T)$ with decreasing temperature might be due to more 
efficient averaging over the disorder associated with a growing 
correlation length. But $\Lambda(T)$ saturates below $\sim$0.2 K and 
there is no evidence for a true phase transition. The relaxation rates are 
broadly distributed ($K \lesssim 0.7$); we are never dealing with 
critical behavior in a homogeneous system. The distribution of rates is 
itself static, at least over the time scale of the experiment 
(${\sim}10~\mu$s), otherwise $\overline{G}(t)$ would tend to be 
averaged to an exponential with the average rate. 

The faster relaxation  (i.e., slower spin fluctuations) for $x = 1.0$ 
than for $x = 1.5$ implies that the relevant critical point 
is near the concentration for which the N\'eel temperature~$T_N$ is 
suppressed to zero ($x \approx 1$). But in spin-glass AgMn $\gamma \to 
1$ from below as $T \to T_g$~\cite{KMCL96}, whereas here 
$\gamma(x{=}1.5) = 0.7,\ \gamma(1.0) = 0.35 < \gamma(1.5)$. A 
mean-field model of a disordered Kondo alloy at a quantum critical 
point~\cite{GrRo99} predicts $\gamma = 1/2$ at $T = 0$, suggestive of 
our results but not in detailed agreement with them.

By definition NFL behavior is a property of the lowest-lying 
excitations of a metal, to which a low-frequency probe such as $\mu$SR 
is extremely sensitive. Many aspects of the spin dynamics in 
UCu$_{5-x}$Pd$_x$ are poorly understood, but the strong disorder, 
glassy behavior, and absence of a phase transition strongly suggest 
that these alloys are quantum spin glasses. 

\smallskip We are grateful to W.~P. Beyermann, A.~H. Castro Neto, and 
L.~P. Pryadko for discussions of these issues, and to A. Amato, C. 
Baines, and D. Herlach for help with the experiments. This research was 
supported in part by the U.S. NSF, Grants DMR-9731361 (U.C. Riverside), 
DMR-9820631 (CSU Los Angeles), and DMR-9705454 (U.C. San Diego), and by 
the Netherlands NWO and FOM, and was performed in part under the 
auspices of the U.S. DOE.

\end{document}